# Optimal design of two-dimensional riser fairings for vortex-induced vibration suppression based on genetic algorithm

Mengyue Chen[a, b], Xiuquan Liu[a,]*, Fulai Liu[a], Min Lou[b]

[a] Centre for Offshore Engineering and Safety Technology, China University of Petroleum (East China), Qingdao, Shandong 266580, China

[b] College of Petroleum Engineering, China University of Petroleum (East China), Qingdao, Shandong 266580, China

## ABSTRACT

Vortex-induced vibration (VIV) manifests its destructiveness on marine risers in deep-water oil and gas development. Many types of riser fairings are employed in the practical applications of ocean engineering due to their effectiveness in suppressing VIV. However, few reports have been found on the optimal design of the fairing profile. In this study, we conduct shape optimization by integrating computational fluid dynamics and genetic algorithm at a low Reynolds number. The fairing profile is parameterized by Bézier curves with control points adopted as design variables, and lift coefficient is selected as the objective function to obtain optimal fairing. The optimal solution shows that the water-drop-shaped fairing presents a high performance in the condition of two degrees of freedom (DOFs), whereas the caudal fin-like-shaped fairing has good stability in 3-DOF condition. Optimization results indicate that the water-drop-shaped fairing suppresses amplitude up to 99% and reduces drag coefficient by almost 51% at $Ur = 5.0$ for the 2-DOF condition. For the 3-DOF condition, the formation of two symmetric vorticities with opposite directions generated at the bilateral of the optimal fairing leads to fairing stability in flow.

## 1. Introduction

Flow past bluff body is a common phenomenon in reality, which may cause vigorous vibration of structures due to the periodic vortex shedding in wake, such as vortex-induced vibration (VIV). VIV is manifested on various engineering branches, such as offshore, civil, and nuclear engineering, and often jeopardizes the lifetime of structures. Since Feng (1968) made a pioneering contribution to the investigation of the VIV response of an elastically mounted cylinder, the theoretical and practical significance of VIV has led to a large amount of analytical, numerical, and experimental studies. Interested readers are referred to review papers by Bearman (1984, 2011), Williamson (1996, 2004, 2008), Sarpkaya (1979, 2004), Ongoren and Rockwell (1988a, 1988b), and Zdravkovich (1996) and books by Blevins (1990), Sarpkaya (2010), and Zdravkovich (1997a, 1997b).

Several methods for reducing VIV have been proposed (Zdravkovich, 1981; Choi et al., 2008;

Rashidi et al., 2016). Specifically, Blevins (1990) summarized four methods, namely, increasing reduced damping, avoiding resonance, streamline cross section, and adding a vortex suppression device. In the fields of ocean and offshore engineering, a vortex suppression device is usually added on a drilling or producing riser to control vortex shedding and VIV. Examples of such device include fairings (Khorasanchi and Huang, 2014; Strandenes et al., 2015; Kristiansen et al., 2015; Constantinides et al., 2015), helical strakes (Vandiver et al., 2006; Korkischko and Meneghini, 2010, 2011; Zhou et al., 2011; Fang et al., 2014; Senga and Larsen, 2017), splitter plates (Bearman, 1965; Shukla et al., 2009; Gu et al., 2012; Huera-Huarte, 2014; Assi et al., 2010, 2015; Lou et al., 2016), and control rods (Wu et al., 2011, 2012; Song et al., 2017).

Among the suppression devices above, fairings are frequently employed in engineering applications due to their low-drag performance. Thus, a great number of researchers have investigated riser fairings. Wang et al. (2015, 2017) studied the VIV and galloping oscillation of a two-dimensional circular cylinder attached with a fixed fairing device by a total variation diminishing approach based upon the elemental velocity vector transformation method. For the VIV of the faired cylinder, the mean drag coefficient could be reduced up to approximately $(10-33)\%$, and the root-mean-square (RMS) lift coefficient could be reduced up to $(30-99)\%$ at Reynolds number ranges from 1,000 to 50,000. The 75°-shaped faired structure might be taken as a proper option considering the influences of attack angles on lift and drag coefficient reduction. For the galloping of the faired cylinder, two factors (i.e., the formation length of vortices in wake and the reattachment of shear layers) seemed to decide whether the galloping oscillation would be prominent. Assi et al. (2011, 2014) conducted experiments to investigate the effect of rotational friction on the stability of a free-to-rotate short-tail fairing. The experiments showed how VIV could be reduced if the rotational friction between cylinder and the short-tail fairing exceeded a critical limit. Yu et al. (2015) investigated the effect of rotational friction on fairing stability on the basis of a numerical study. They introduced fictitious methods and successfully stabilized simulations by avoiding the so-called added mass effect. The simulation results showed that friction coefficient varied depending on Reynolds numbers.

For the research of the profile shape or geometric configuration of riser fairings in the past decade, Pontaza et al. (2012) evaluated a U-shaped fairing on the basis of numerical simulations. The analysis results showed that the fairing design was effective at suppressing VIV and yielded a low drag coefficient, and its latching mechanism was adequate for use in calm-sea states with four-knot current speeds. Xie et al. (2015) simulated three-dimensional free-to-rotate U-shaped fairings with three different geometric configurations (i.e., a homogeneous one without a gap, a single gap with two adjacent fairings, and two gaps separating three adjacent fairings) in a cross flow at Reynolds number ranges from 100 to 10,000. The effect of the gap increased with the Reynolds number, and the effect of friction in the fairing rotation was relatively small, which differed from the result obtained in previous studies. Law et al. (2016, 2017) presented a set of VIV simulations for short-crab claw (SCC) fairings with different lengths. The SCC fairing with a long length performed efficiently by suppressing amplitude up to 84% and reduced drag coefficient by 40%, which implied that it lowered the influences of vortex interactions and led to VIV suppression by offsetting the shedding away of vortices from the main cylinder. The authors also proposed a new design of fairing, namely, "hinged C-shaped" or "connected-C." The numerical simulation results showed that the new fairing design was efficient with respect to VIV suppression and drag reduction. A pair of counter-rotating circulations were generated at the bilateral of a device and stabilized the device in the flow. Baarholm et al. (2015) conducted experiments to investigate the VIV responses of fairings in different profiles. Several profiles showed clear capability to flutter, whereas other fairings were stable over the entire velocity range and range of inflow angles tested.

From the previous research above, we can conclude that the riser fairing profile plays a key role in VIV reduction. However, the previous research merely focused on the effect of VIV suppression of different fairing types, such as U-shaped, water-drop-shaped, and SCC fairing. The research of shape optimization of riser fairings has received little attention. Therefore, in the present study, shape optimization is conducted numerically by integrating computational fluid dynamics (CFD) and genetic algorithm (GA) to obtain a high-performance riser fairing. The riser fairing profile is parameterized by Bézier curves, the control points of which are used as design variables, and lift coefficient is selected as the objective function to obtain optimal fairing profile. We initially conduct VIV simulation with a plain cylinder for validation and comparison. The shape optimization of fairing is then implemented at low Reynolds number with different inflow steam velocities and numbers of design variables. The fairing condition in two degrees of freedom (DOFs) or 3-DOFs is evaluated.

| **Nomenclature** | |
|---|---|
| $Re$ | Reynolds number |
| $p$ | Fluid pressure |
| $\rho$ | Fluid density |
| $U$ | Fluid velocity |
| $Ur$ | Reduced velocity |
| $D$ | Diameter of the cylinder |
| $A$ | Transverse amplitude |
| $m$ | Mass of structure |
| $m^*$ | Mass ratio |
| $k$ | Stiffness of the structure |
| $c$ | Damping of the structure |
| $\xi$ | Damping ratio |
| $C_D$ | Drag coefficient |
| $C_L$ | Lift coefficient |
| $F_D$ | Drag force |
| $F_L$ | Lift force |
| $f_n$ | Natural frequency of the structure |
| $F_n$ | Reduced natural frequency of the structure |

## 2. Numerical models

### 2.1 Governing equations of fluid and structure motion

The incompressible Navier–Stokes equations are employed in the fluid model and solved by the finite element method. The arbitrary Lagrangian–Eulerian method is used in this study to simulate the VIV of the structure. In two dimensions, the dimensionless governing equations are expressed as follows:

$$\frac{\partial u}{\partial x} + \frac{\partial v}{\partial y} = 0, \tag{1}$$

$$\frac{\partial u}{\partial t} + u\frac{\partial u}{\partial x} + v\frac{\partial u}{\partial y} = -\frac{\partial p}{\partial x} + \frac{1}{Re}\left(\frac{\partial^2 u}{\partial x^2} + \frac{\partial^2 u}{\partial y^2}\right), \tag{2}$$

$$\frac{\partial v}{\partial t} + u\frac{\partial v}{\partial x} + v\frac{\partial v}{\partial y} = -\frac{\partial p}{\partial y} + \frac{1}{Re}\left(\frac{\partial^2 v}{\partial x^2} + \frac{\partial^2 v}{\partial y^2}\right), \tag{3}$$

where $x$ and $y$ are the Cartesian coordinates in the streamwise (parallel to the flow) and transverse (normal to the flow) directions of the flow, respectively; $u$ and $v$ are the fluid velocity components along the $x$- and $y$-coordinates, respectively; and $p$ represents pressure.

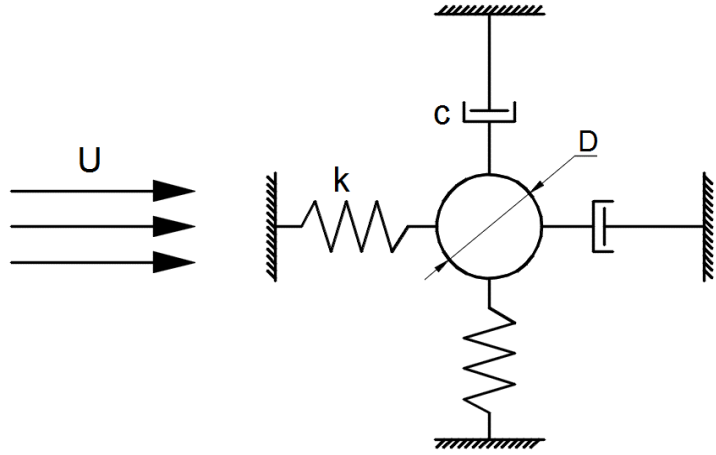

**Fig. 1.** Schematic model of the oscillating structure.

A schematic model of the oscillating structure is presented in Fig. 1. The governing equations of oscillator motion in the two directions are written as follows:

$$\ddot{X} + 4\pi F_n \xi \dot{X} + (2\pi F_n)^2 X = \frac{2C_D}{\pi m^*}, \tag{4}$$

$$\ddot{Y} + 4\pi F_n \xi \dot{Y} + (2\pi F_n)^2 Y = \frac{2C_L}{\pi m^*}, \tag{5}$$

where $\ddot{X}$, $\dot{X}$, and $X$ are the acceleration, velocity, and displacement of the oscillator in the streamwise direction of the flow, respectively; $\ddot{Y}$, $\dot{Y}$, and $Y$ denote the same quantities associated with the transverse motion; $F_n = f_n D / U$ is the reduced natural frequency of the structure; $\xi = c / (2\sqrt{km})$ is the structural damping ratio; $m^* = 4m / \rho \pi D^2$ is the dimensionless mass of the structure; and $C_D = F_D / (\rho D U^2 / 2)$ and $C_L = F_L / (\rho D U^2 / 2)$ are the drag and lift coefficients, respectively. $m$ is the actual mass of the structure, excluding added mass. For numerical simulation, the effect of added mass is considered and calculated through Navier–Stokes equations directly (Sarpkaya, 2004). The above governing equations that describe fluid and structure motion are solved within the COMSOL Multiphysics.

## 2.2 Grid generation and validation

The parameters of plain cylinder in the present study are the same as in the numerical experiment of Zhao et al. (2014), with $D = 0.003$ m, $m^* = 2$, $k = 5.023$ N / m, and $f_n = 3$ Hz. The structural damping ratio $\xi$ is set to zero to encourage high-amplitude oscillations. The calculation area should be sufficiently large to eliminate the effect of blockage on flow past a vibrating bluff body. The width of the computational domain is set to $20D$ and the length is set to $30D$ according to the conclusions from Prasanth et al. (2006).

**Table 1**
Summary of the lift and drag coefficients and amplitude response of the cylinder for five different finite element meshes.

| Mesh | Element | RMS $C_L$ | Mean $C_D$ | Max (A/D) |
|---|---|---|---|---|
| 1 | 7211 | 0.1372 | 2.0947 | 0.5852 |
| 2 | 4713 | 0.2116 | 2.1643 | 0.6042 |
| 3 | 3394 | 0.2524 | 2.2025 | 0.6252 |
| 4 | 2304 | 0.2760 | 2.2216 | 0.6315 |
| 5 | 1635 | 0.2837 | 2.2424 | 0.6281 |

An appropriate mesh size and an adequate number of elements are essential to simulate the fluid–solid interaction problem successfully. However, an excessive mesh is unconducive, especially considering the long process of shape optimization. Thus, we investigate several cases for different finite element meshes. All the results are summarized in Table 1. The mesh with excessive elements or extremely few elements cannot simulate the VIV of plain cylinder correctly. In view of the shape change in the process of shape optimization, we select the mesh size conservatively (i.e., mesh 3), and the mesh is illustrated in Fig. 2.

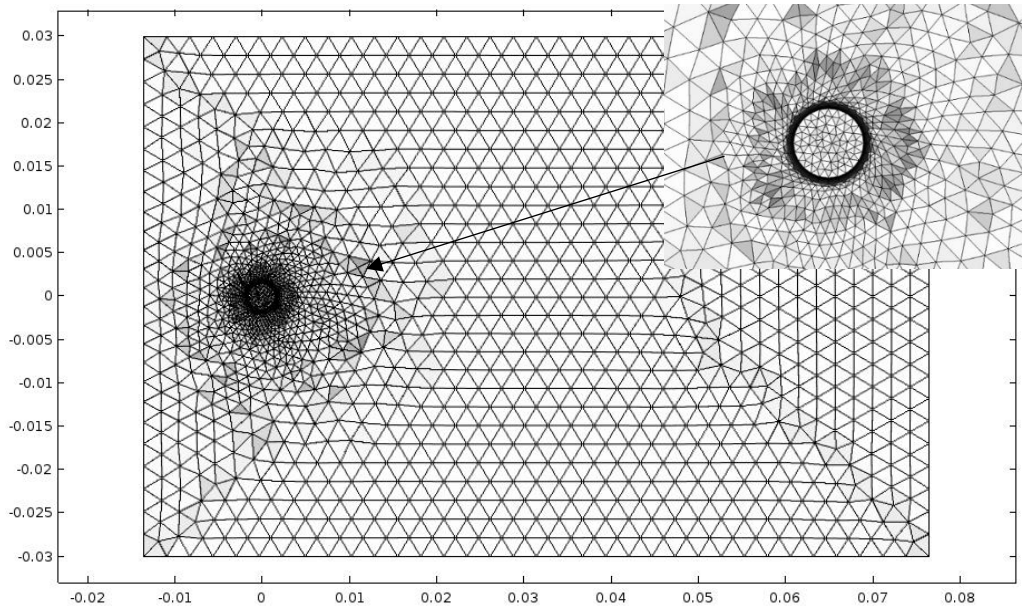

**Fig. 2.** Finite element mesh around the bare cylinder.

The simulation environment should be validated with available data from published literature before implementing the study of shape optimization. Here, we compare the VIV simulation results with the numerical simulation results of Zhao et al. (2014). The drag coefficients and the amplitude of in-line direction are not compared because the drag coefficients are extremely small and relative errors are amplified. The results are summarized and illustrated in Figs. 3 and 4. These figures show a good agreement between the present and previous works.

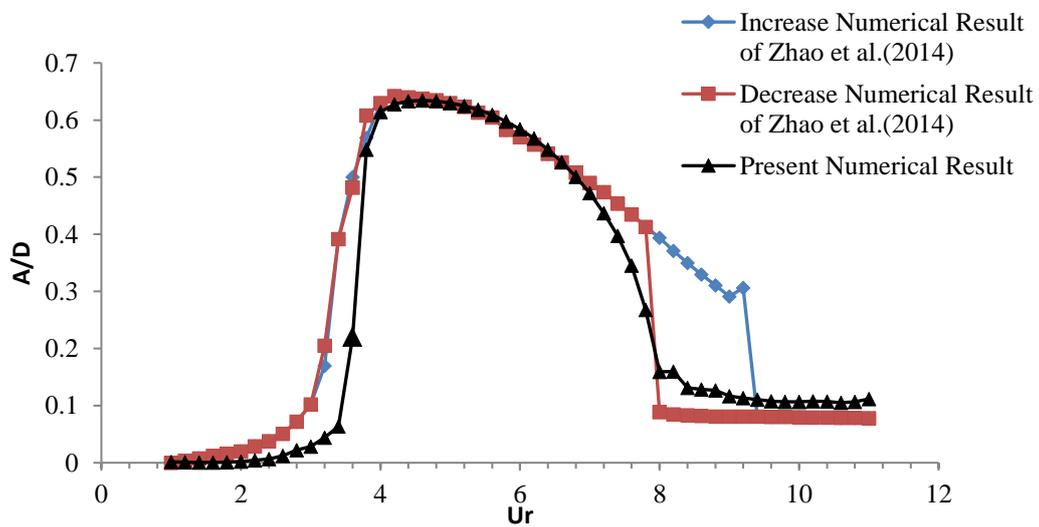

**Fig. 3.** Comparison of the VIV oscillation amplitudes of the structure displacement in cross-flow direction.

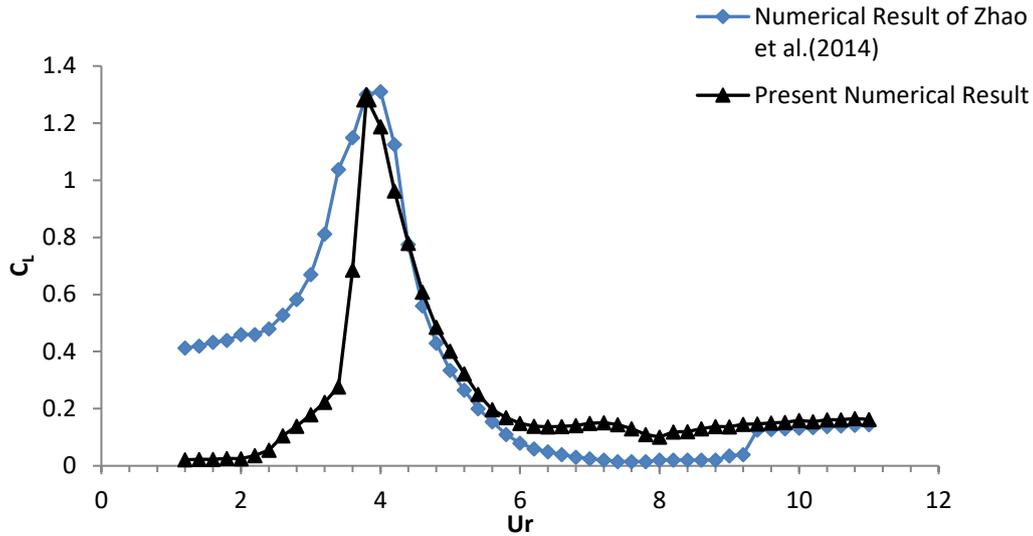

**Fig. 4.** Comparison of the lift coefficients of the VIV oscillation of the structures.

## 3. Shape optimization

Optimization is the process of finding the optimal solution while satisfying specific constraints by minimizing or maximizing the objective function. Many optimization algorithms have been presented over the past several decades due to their own features. In this study, we aim to determine a distinct shape that differs from the current fairing and explore the mechanism of what fairing type has a good performance. Thus, GA is selected as the optimization algorithm for the feature of global optimum and the mechanism of natural selection. Since GA was invented by Holland in the 1960s, it has demonstrated its effectiveness and maturity in many application fields. The basic processes of GA are specified as follows (Mitchell, 1996):

1. Randomly generated a population of $n$ chromosomes.
2. Calculate the fitness function of each chromosome in the population.
3. Repeat the following steps until $n$ offspring is created:
    a) Select a pair of parent chromosomes from the current population.
    b) With crossover probability (crossover rate), crossover the pair at a randomly selected point to form two offspring.
    c) Mutate the two offspring at each locus with mutation probability (mutation rate), and place the resulting chromosomes in the new population.
4. Replace the current population with the new population.
5. Return to Step 2.

Several characteristics, such as objective function, design variable, and constraint, should be prescribed before implementing optimization. Table 2 lists the optimization parameters used in this study.

Table 2

Parameters for GA.

|  | Values |
|---|---|
| Population | 25 |
| Generation | 20 |
| Crossover fraction | 0.8 |
| Migration fraction | 0.2 |

## 3.1 Objective function

A properly optimized riser fairing will be stable even in the circumstance that VIV is likely to occur. Lift coefficient is usually used to evaluate the stability of VIV problems. The vibration amplitude of in-line direction is relatively small for a riser fairing, and lift coefficient is a dimensionless parameter. However, in view of the shape deformation during the optimization, the classic lift coefficient is infeasible in the present work for two reasons. First, the hydraulic diameter of the object during optimization is varying. Second, for ocean engineering application, we prefer to focus on the total force of riser in cross-flow direction. Therefore, we redefine lift coefficient by considering the diameter of the marine riser (i.e., the diameter of the cylinder $D$) as the hydraulic diameter, and the objective function is to minimize the lift coefficient $C_L$.

## 3.2 Design variables

Selecting a proper method for representing the riser fairing shape is essential. As a frequently adopted method in the shape optimization of airfoils, the Bézier curve has the advantage of easy control of the airfoil profiles with a small number of points, which leads to a smooth change in the airfoil shapes (Park and Lee, 2010). Thus, the fairing shape is parameterized by Bézier curves, the control points of which are used as the design variables. The rational Bézier curve is expressed as follows:

$$B(t) = \frac{\sum_{i=0}^{n} b_{i,n}(t) P_i \omega_i}{\sum_{i=0}^{n} b_{i,n}(t) \omega_i} \quad t \in [0,1], \tag{6}$$

$$b_{i,n}(t) = \binom{n}{i} t^i (1-t)^{n-i}, \tag{7}$$

where $b_{i,n}$ is a blending function; parameter $t$ is bounded from 0 to 1; $P_i$ refers to the coordinates of the control point; and $\omega_i$ denotes the weights.

We divide the design variables into three cases (i.e., Cases A, B, and C). Fig. 5 illustrates the three different schematics of the riser fairing using Bézier curves. Case 1 (Fig. 5a) comprises only one Bézier curve ($\overline{AC}$), two control points ($B, C$), and three coordinate variables ($v_1 - v_3$). Case 2 (Fig. 5b) includes two Bézier curves ($\overline{AC}, \overline{CE}$), four control points ($B - E$), and seven coordinate variables ($v_1 - v_7$). Case 3 (Fig. 5c) consists of three Bézier curves ($\overline{AC}, \overline{CE}, \overline{EG}$), six control points ($B - G$), and eleven coordinate variables ($v_1 - v_{11}$). The fairing profile is a symmetrical shape along the *x*-axis.

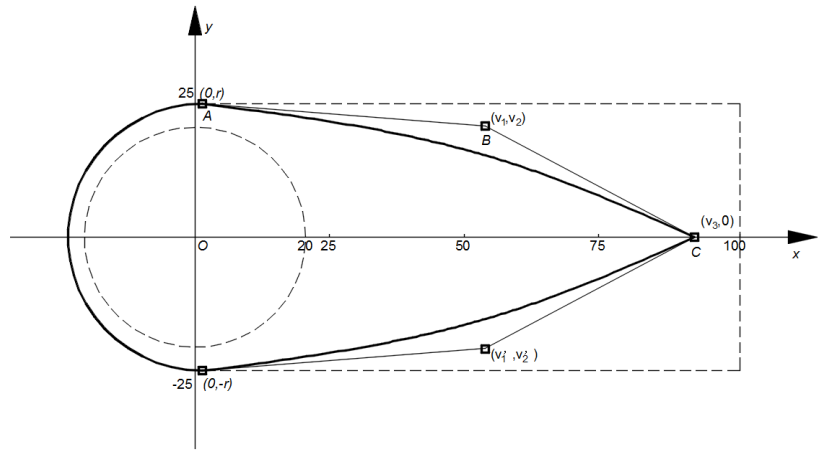

(a) Case A

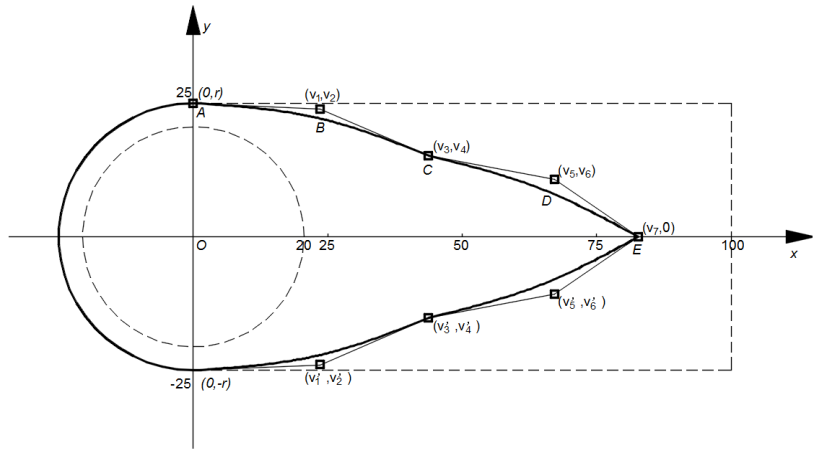

(b) Case B

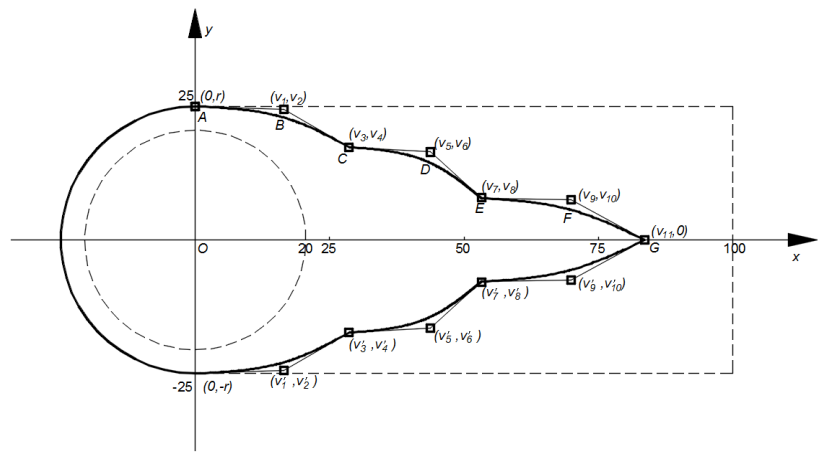

(c) Case C

**Fig. 5.** Schematics of the riser fairing for different cases.

## 3.3 Constraints

For each case, the side constraint is a rectangular area of $2D$ long and $0.5D$ wide. This rectangular area is separated as $100 \times 25$ blocks to save time of shape optimization. In light of the riser fairing in ocean engineering application and installation, the distance between each control point to the origin of coordinates is greater than 20 units. Control points $C$ (Fig. 5a), $E$ (Fig. 5b), and $G$ (Fig. 5c) are changed randomly along the $x$-axis. In reality, a fairing is a free-to-rotate device attached to the riser. The object in Case C (Fig. 5c) is rotatable to satisfy the practical necessity, and the rotational damping is set to zero to encourage high-amplitude oscillations.

## 4. Results and discussion

Despite the obvious difference between simulation and physical environments, the dynamics and response characteristics of VIV have their roots in two-dimensional low Reynolds flow (Law et al., 2017). Thus, an investigation at low Reynolds number is worth to conduct. We present our results in the following two subsections to validate the effectiveness of shape optimization on VIV suppression. In Section 4.1, we investigate Cases A and B in different reduced velocities to evaluate the influences of the number and location of design parameters on VIV suppression and analyze the influences of different reduced velocities on shape optimization. As previously mentioned, transverse and streamwise 2-DOF motions are considered for the investigation. In Section 4.2, we investigate Case C and extend our research to 3-DOF motion, in which rotational motion is considered. The characteristic VIV response of an optimal fairing shape in two specific reduced velocities is analyzed with emphasis on vorticity dynamics.

### 4.1 Assessment for 2-DOF

Fluid velocity, which is relevant to the amplitude and force of structures, plays a key role in VIV response. Figs. 3 and 4 illustrate the influence of fluid velocity at low Reynolds number. We select fluid velocities as $Ur = 3.0$, $Ur = 5.0$, and $Ur = 7.0$, which are representative for initial and lower branches, to further investigate the influences of different reduced velocities on shape optimization. The number of design parameters is directly related to the optimization. Therefore, we compare Cases A (three coordinate variables) and B (seven coordinate variables) to determine the effect of the number of design parameters on shape optimization. The dimensionless VIV parameters are similar to those in the study of Zhao et al. (2014), where the mass ratio $m^* = 2$, the stiffness $k = 5.023$ N / m, and the Reynolds number $Re \in [100,400]$.

#### 4.1.1 Optimal shape

Figs. 6 and 7 depict all the optimal solutions for shape optimization in Cases A and B, respectively. Tables 3 and 4 present the geometric parameters of optimal shape. The figures illustrate that the optimal shape of fairing at $Ur = 3.0$ is narrow in the middle, whereas the optimal shape of fairing at $Ur = 7.0$ is wide in the middle. Despite the interesting phenomenon found in Cases A and B, we cannot draw a conclusion rashly that the optimal shape of fairing becomes wide as velocity increases. In fact, the computational cost of shape optimization limits the optimization for a continuous range of velocities; nevertheless, several discrete velocities can be selected to implement the investigation. The stochasticity of GA also affects the optimal solution to some degree. During the research, we actually acquire some strange optimal shapes after optimization, which are far from the optimal shapes in Figs. 6 and 7.

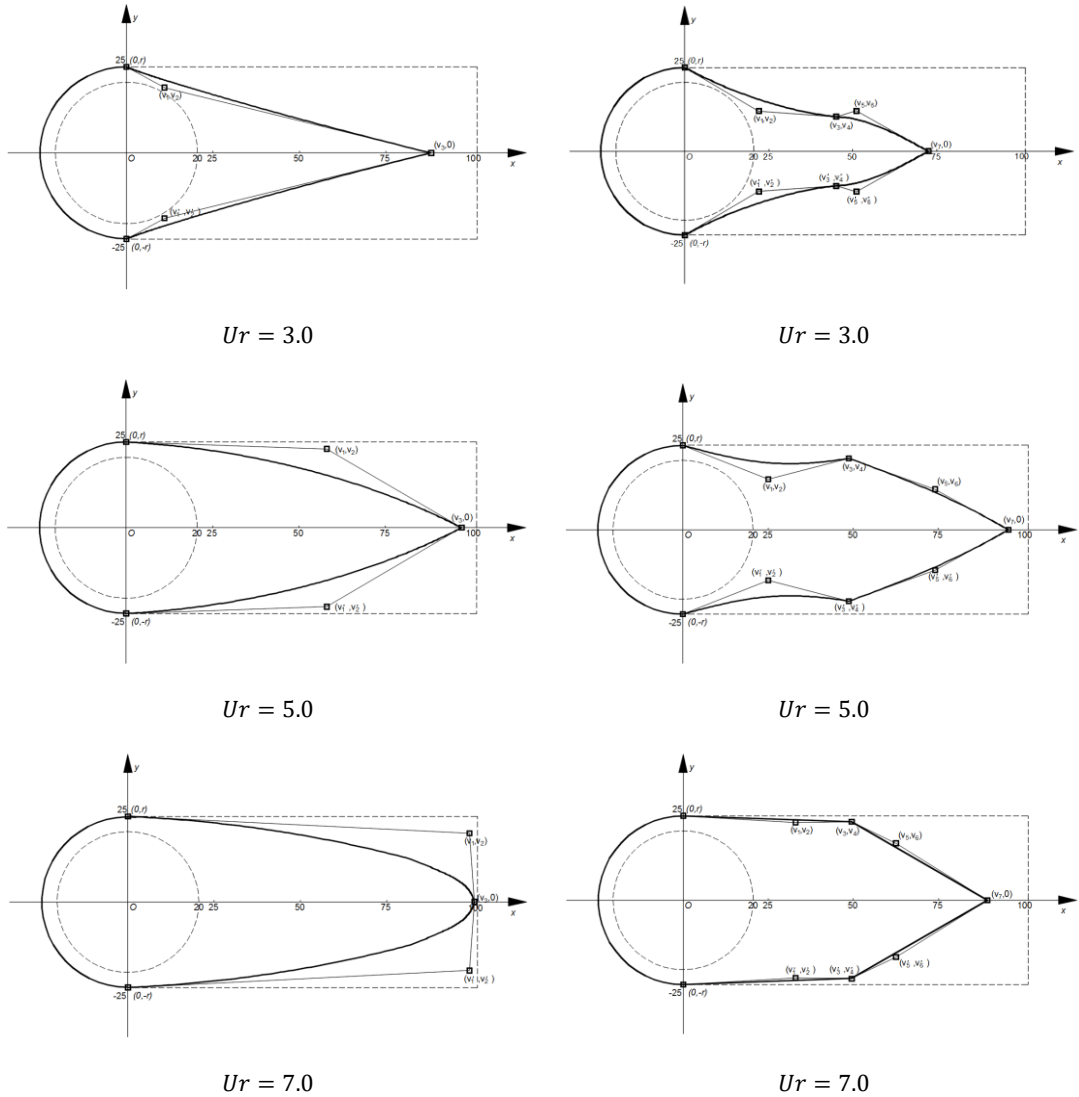

$Ur = 3.0$            $Ur = 3.0$

$Ur = 5.0$            $Ur = 5.0$

$Ur = 7.0$            $Ur = 7.0$

**Fig. 6.** Optimal shape of Case A.      **Fig. 7.** Optimal shape of Case B.

**Table 3**
Geometric parameters of optimal shape for Case A.

| Reduced velocity | Coordinate value | |
|---|---|---|
| $Ur = 3.0$ | (0.022,0.038) | (0.176,0) |
| $Ur = 5.0$ | (0.116,0.046) | (0.194,0) |
| $Ur = 7.0$ | (0.198,0.040) | (0.198,0) |

**Table 4**
Geometric parameters of optimal shape for Case B.

| Reduced velocity | Coordinate value | | | |
|---|---|---|---|---|
| $Ur = 3.0$ | (0.044,0.024) | (0.090,0.020) | (0.102,0.024) | (0.144,0) |
| $Ur = 5.0$ | (0.050,0.030) | (0.096,0.042) | (0.148,0.024) | (0.190,0) |
| $Ur = 7.0$ | (0.066,0.046) | (0.098,0.046) | (0.124,0.032) | (0.178,0) |

### 4.1.2 Amplitude response

The transverse amplitude response data of Case A, Case B, and plain cylinder are summarized in Table 5. For the three different reduced velocities, the devices of Cases A and B both suppress the VIV effectively. When the plain cylinder starts to vibrate slightly at $Ur = 3.0$, the amplitude is reduced by 99.978% for Case A and by 99.933% for Case B. When the plain cylinder experiences large amplitudes at $Ur = 5.0$ and $Ur = 7.0$, the amplitudes for Case A are reduced by 99.997% and 98.560% and those for Case B are reduced by 99.365% and 99.899%, respectively. The amplitude response shows that in the 2-DOF condition, the difference between Cases A and B is relatively small. Unlike fluid velocity, the number of design parameters plays a less significant role in shape optimization.

Table 5
Transverse reduced amplitude.

| | $Ur = 3.0$ | $Ur = 5.0$ | $Ur = 7.0$ |
|---|---|---|---|
| Plain cylinder | 2.84E-02 | 6.30E-01 | 4.72E-01 |
| Case A | 6.27E-06 | 1.64E-05 | 6.80E-03 |
| Case B | 1.91E-05 | 4.00E-03 | 4.77E-04 |

### 4.1.3 Lift and drag coefficients

We consider lift and drag coefficients to further analyze the stability of different cases in different fluid velocities. The data of RMS lift coefficient $C_L$ and mean drag coefficient $C_D$ are summarized in Tables 6 and 7, respectively. Lift coefficient is reduced by 99.915% for $Ur = 3.0$, 99.923% for $Ur = 5.0$, and 78.870% for $Ur = 7.0$ in Case A and by 97.819% for $Ur = 3.0$, 99.754% for $Ur = 5.0$, and 97.241% for $Ur = 7.0$ in Case B. Tables 5 and 6 imply that the variances between reduced amplitude and lift coefficient of Cases A and B are similar.

Table 6
RMS lift coefficient.

| | $Ur = 3.0$ | $Ur = 5.0$ | $Ur = 7.0$ |
|---|---|---|---|
| Plain cylinder | 1.79E-01 | 3.95E-01 | 1.49E-01 |
| Case A | 1.52E-04 | 3.05E-04 | 3.14E-02 |
| Case B | 3.90E-03 | 9.71E-04 | 4.10E-03 |

Drag coefficient is reduced to 93.747% only in Cases A and B due to its slight vibration at $Ur = 3.0$. However, at $Ur = 5.0$, the drag coefficient is reduced to 50.002% for Case A and 48.570% for Case B; at $Ur = 7.0$, the drag coefficient is reduced to 62.499% for Case A and 65.625% for

Case B. Although the drag coefficient is not part of the objective function in shape optimization, the performances of Cases A and B are satisfied in terms of drag suppression. Table 7 also shows that the number of design parameters plays a less important role than fluid velocity in shape optimization.

**Table 7**
Mean drag coefficient.

|  | $Ur = 3.0$ | $Ur = 5.0$ | $Ur = 7.0$ |
|---|---|---|---|
| Plain cylinder | 1.4632 | 2.3045 | 1.6125 |
| Case A | 1.3717 | 1.1523 | 1.0078 |
| Case B | 1.3717 | 1.1193 | 1.0582 |

To summarize, in the 2-DOF cases, fluid velocity plays a significant role in shape optimization. The optimal shapes in different velocities are distinct from one another. The number of design parameters plays a less significant role than fluid velocity in shape optimization; however, the optimal shapes in different numbers of coordinate variables are similar to one another. The optimal shape in both cases is water drop shape or similar to water drop shape. The high performance of the optimal fairing may explain the minimal effect of the number of design parameters. The performance of the water-drop-shaped fairing is sufficient for the 2-DOF condition. Other shape does not need to evolve. All the optimal shapes have high performances in VIV suppression in terms of transverse amplitude and lift and drag coefficients.

## 4.2 Assessment for 3-DOF

Assi et al. (2009) mentioned that the instability of free-to-rotate suppressors is directly related to the level of rotational resistance encountered in the system, as well as geometric parameters. We verify this phenomenon in the present work and find that the optimal shape in Case A or B is unsuitable for the 3-DOF condition. In other words, when the optimal fairing is rotatable and the rotational damping is set to zero, the suppressor may induce vigorous oscillations excited by galloping or fluttering at specific velocities. We focus on the improvement of geometric parameters to determine a stable free-to-rotate suppressor. An appropriate rotational resistance can conduce to the stability of free-to-rotate suppressors. The additional computing resource for rotational damping is inefficient in the process of shape optimization. Therefore, in this section, we extend the number of design parameters to 11 and select fluid velocity $Ur = 4.0$, such that the amplitude and force in cross-flow direction reach the maximum. The fluid velocity $Ur = 5.5$ is also considered for comparison. The dimensionless VIV parameters in Case C are similar to those in Cases A and B.

### 4.2.1 Optimal shape

Fig. 8 shows the two optimal solutions for shape optimization in the assessment for 3-DOF. Some interesting optimal shapes are acquired during the optimization. As illustrated in Fig. 8, the fairing tip presents a caudal fin-like shape, which has a high performance under the condition of no torsional resistance. For convenience, we name the two optimal shapes as Caudal Fin-like Shapes A and B. Caudal Fin-like Shapes A and B are the optimal shapes at $Ur = 4.0$ and $Ur = 5.5$, respectively. The geometric parameters are provided in Table 8. Caudal Fin-like Shape A is considerably wider in the middle than Caudal Fin-like Shape B. We evaluate the aspects of amplitude response, lift and drag coefficients, pressure distributions, and vortex patterns to further compare the effectiveness of the two different optimal shapes.

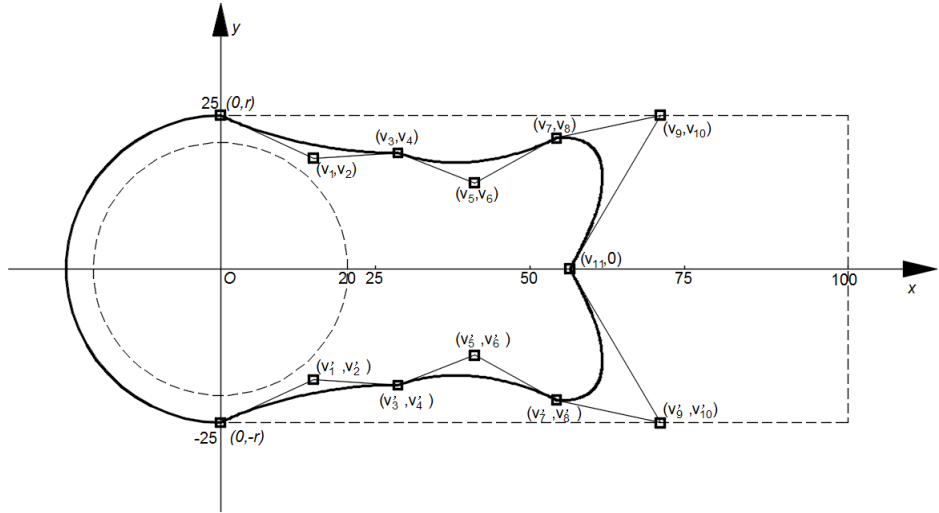

Caudal Fin-like Shape A

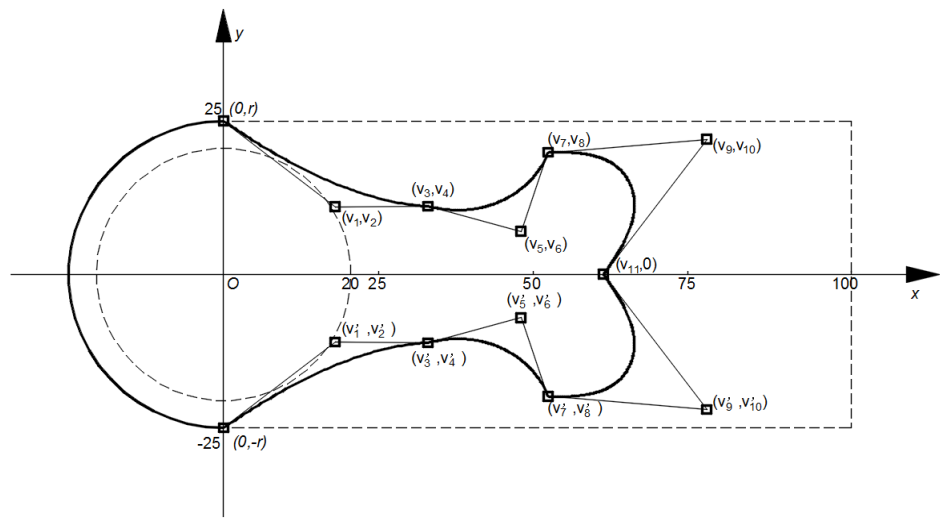

Caudal Fin-like Shape B

**Fig. 8.** Optimal shapes of Case C.

**Table 8**
Geometric parameters of optimal shapes for Case C.

| Reduced velocity | Coordinate value | | |
|---|---|---|---|
| Caudal Fin-like Shape A | (0.030,0.036) | (0.056,0.038) | (0.082,0.028) |
| $Ur = 4.0$ | (0.106,0.042) | (0.142,0.050) | (0.112,0) |
| Caudal Fin-like Shape B | (0.044,0.032) | (0.068,0.034) | (0.088,0.022) |
| $Ur = 5.5$ | (0.118,0.048) | (0.158,0.040) | (0.124,0) |

**4.2.2 Amplitude response**

Unlike the analysis of amplitude response in the above assessment, we investigate the two optimal shapes at range of reduced velocity $Ur \in [1,10]$ to cover the pre-lock-in, lock-in, and post-lock-in regions of vortex synchronization. Fig. 9 illustrates the amplitude curves of Caudal Fin-like Shape A, Caudal Fin-like Shape B, and plain cylinder. Both Caudal Fin-like Shapes A and B suppress the VIV effectively at particular ranges. At range of reduced velocity $Ur \in [2,5]$, the performance of Caudal Fin-like Shape A is better than that of Caudal Fin-like Shape B. The amplitude of Caudal Fin-like Shape A is reduced to 47.377% at $Ur = 4.0$. By contrast, the performance of Caudal Fin-like Shape B is better than that of Caudal Fin-like Shape A at the range of $Ur \in [5,10]$. The amplitude of Caudal Fin-like Shape B is reduced to 8.437% at $Ur = 6.0$. This phenomenon indicates that the range of high performance for an optimal shape is limited. Caudal Fin-like Shape B starts to vibrate vigorously after $Ur > 6.0$. This phenomenon has also been found in Caudal Fin-like Shape A at the range of $Ur \in [5, \infty]$. The vibrations are attributed to the galloping oscillations of the optimal fairing. The same observation has also been reported in previous studies (Assi et al., 2009). Therefore, the VIV performance of optimal shape is worse than that of plain cylinder after $Ur > 7.5$.

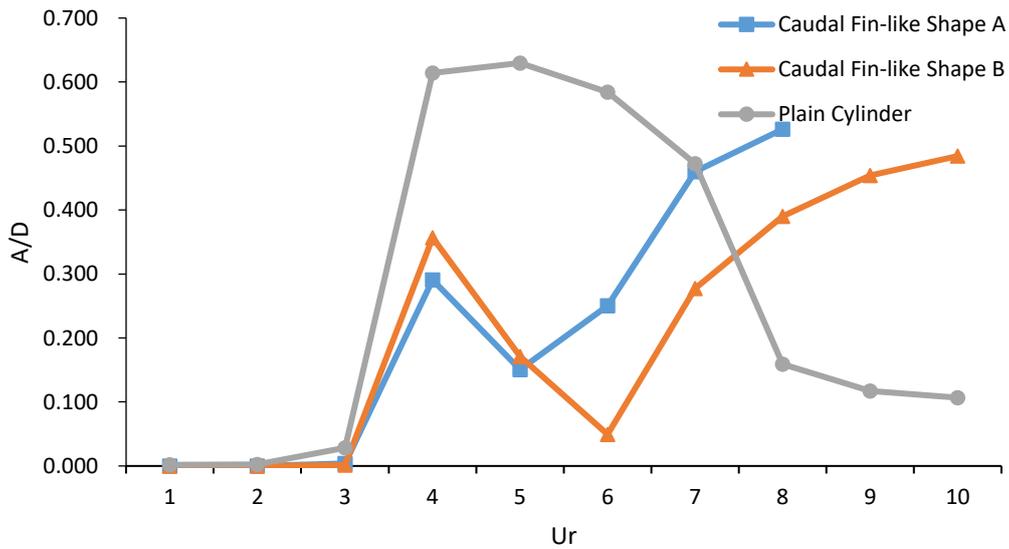

**Fig. 9.** Variation in response amplitude with reduced velocity.

**4.2.3 Lift and drag coefficients**

We analyze RMS lift and mean drag coefficients to validate the stability of different optimal shapes at range of reduced velocity $Ur \in [1,10]$. The curves of lift and drag coefficients are illustrated in Figs. 10 and 11, respectively. For lift coefficient, Caudal Fin-like Shapes A and B play an effective role at the range of $Ur \in [1,6]$. At $Ur = 4.0$, lift coefficient is reduced by 96.384% for Caudal Fin-like Shape A; at $Ur = 6.0$, it is reduced by 48.987% for Caudal Fin-like Shape B.

Fig. 10 demonstrates that the sharp peak in the lift coefficient for plain cylinder and Caudal Fin-like Shape B occurs at $Ur = 4.0$, which indicates the transition from the initial branch to the lower branch. However, the sharp peak for Caudal Fin-like Shape A is unclear. The maximum lift coefficient for Caudal Fin-like Shape A is at $Ur = 7.0$, which is ascribed to the galloping oscillations according to Fig. 9.

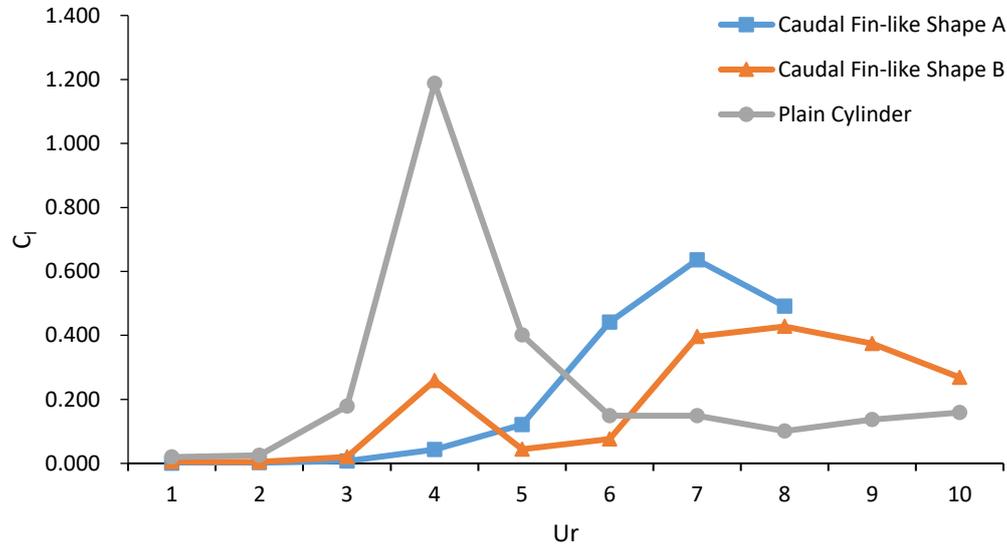

**Fig. 10.** Variation in lift coefficient with reduced velocity.

Drag coefficient is also reduced in both shapes. Fig. 11 depicts that the performances of Caudal Fin-like Shapes A and B are similar to each other. At $Ur = 4.0$, drag coefficient is reduced to 78.108% for Caudal Fin-like Shape A and to 79.617% for Caudal Fin-like Shape B; at $Ur = 6.0$, it is reduced to 61.905% for Caudal Fin-like Shape A and to 60.287% for Caudal Fin-like Shape B. Drag coefficient is not part of the objective function in shape optimization; therefore, the performances of both optimal shapes are poor compared with those in previous studies (Law et al., 2017) but are satisfied in terms of drag suppression. The VIV simulation of Caudal Fin-like Shape A after $Ur > 8.0$ is difficult to conduct in the same simulation environment of Caudal Fin-like Shape B or other previous investigation. For better comparison, we dismiss the data of Caudal Fin-like Shape A after $Ur > 8.0$ in the section of amplitude response and lift and drag coefficients.

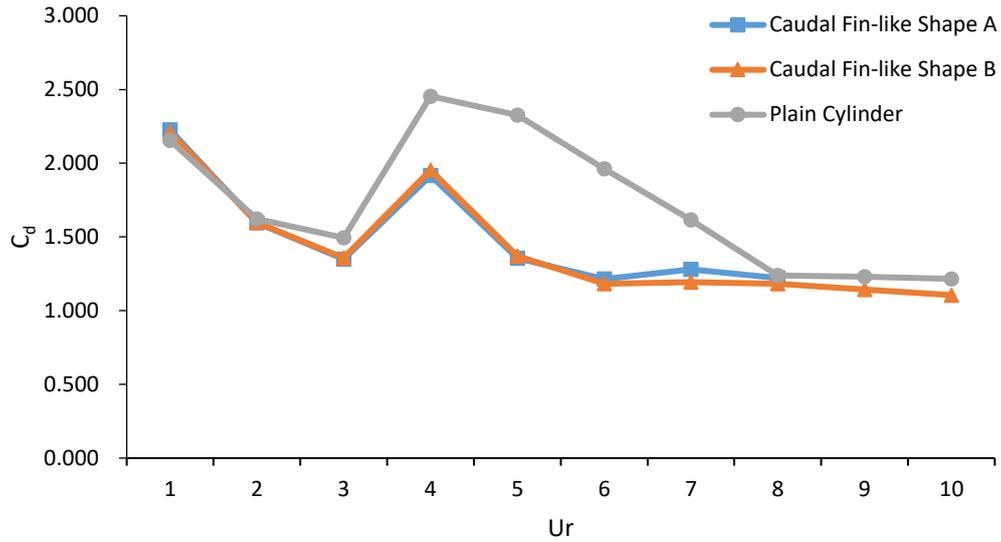

**Fig. 11.** Variation in drag coefficient with reduced velocity.

### 4.2.4 Vortex patterns

Unlike the above data of amplitude and lift and drag coefficients, the flow fields in terms of instantaneous vorticities, streamlines, and pressure fields are visualized to determine the VIV responses of the two optimal fairings. In this subsubsection, we initially compare the fluid fields between the plain cylinder and Caudal Fin-like Shape A to distinguish the VIV generation mechanisms of these structures. We then elucidate the reasons for VIV suppression for Caudal Fin-like Shape B through elaborate instantaneous flow visualizations of streamlines and vorticity contours.

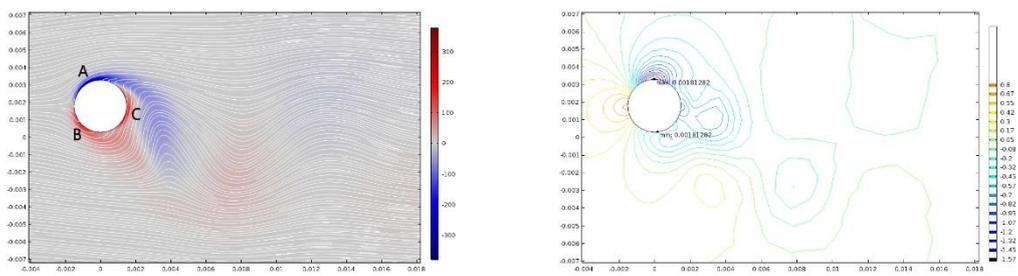
(a) t = 1/4T

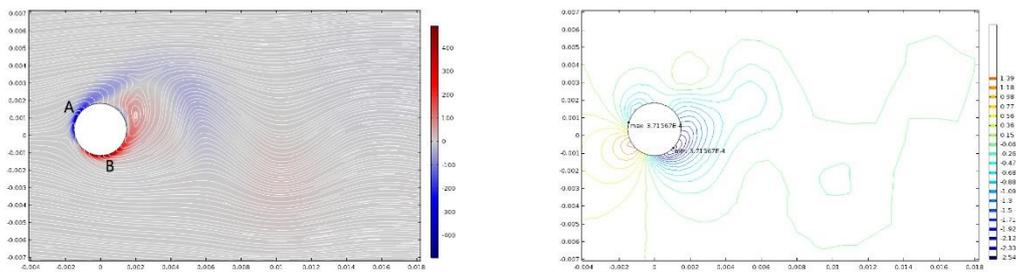
(b) t = 2/4T

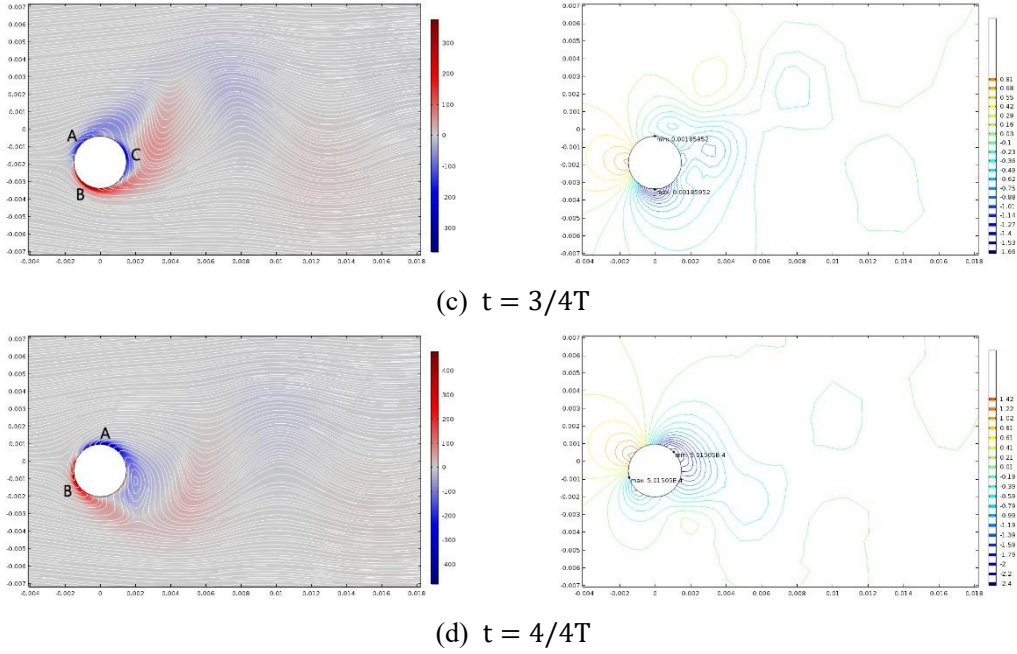

(c) t = 3/4T

(d) t = 4/4T

**Fig. 12.** Instantaneous fluid fields in terms of vorticity contour, streamline trace, and pressure distribution for plain cylinder over one period of VIV oscillation cycle.

Reduced velocity $Ur = 4.0$, with maximum amplitude, is selected to conduct the comparative analysis according to previous studies from other scholars and the discussion in the above section. Figs. 12 and 13 depict the flow fields of the plain cylinder and Caudal Fin-like Shape A during one period of vibration, respectively; vorticity contour, streamline trace, and pressure distribution are included. The classical terminology proposed by Williamson and Roshko (1988) to identify vortex-shedding patterns (e.g., 2S, 2P, P+S) is adopted for better elucidation. The cycle period that we discuss in the present work commences at the zenith of oscillation amplitude. Instantaneous vorticity contours with superimposed streamlines (left) and the pressure distribution with extreme point (right) are shown at the interval of one quarter period.

Fig. 12 presents that the flow for plain cylinder shows a 2S mode of vortex shedding, in which two single Karman vortices with reversed rotation directions shed alternately from the rear of the plain cylinder during each vortex-shedding period. The shear layer separates from the top or bottom of the plain cylinder cyclically. The specific separate point varies obviously depending on the location of the plain cylinder. The varying location of the minimum pressure distribution conduces to the identification of the separate point. Three separate points (i.e., points A, B, and C) are observed without generating vorticity when the plain cylinder reaches the maximum or minimum of oscillation amplitude. The magnitudes of vorticity and pressure show that the fluid field is relatively intense and induces vigorous vibration of the plain cylinder according to different shades. A similar phenomenon has been observed by many scholars in previous experimental or numerical studies for the VIV response of an elastically mounted cylinder in laminar or turbulence fluid.

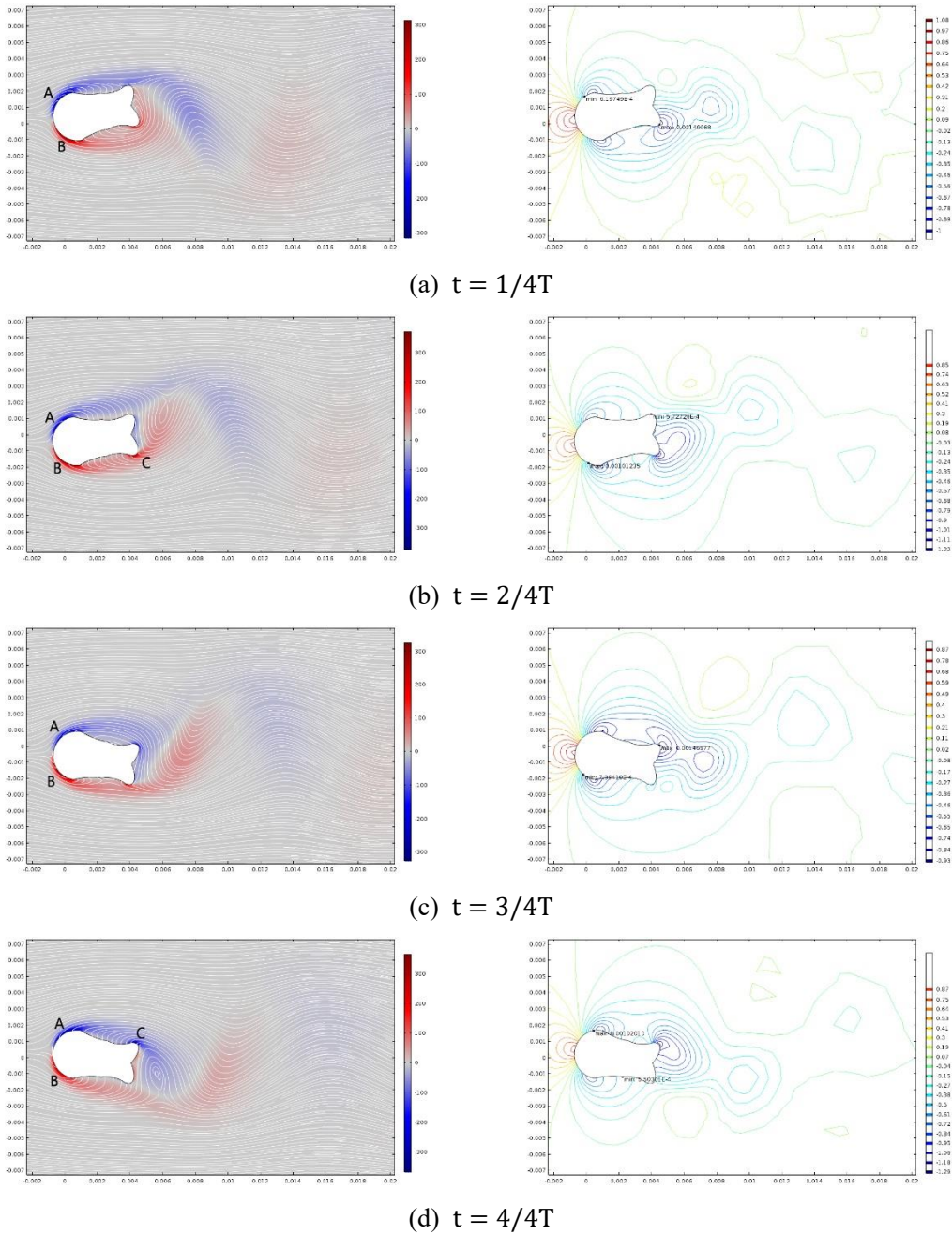

(a) t = 1/4T

(b) t = 2/4T

(c) t = 3/4T

(d) t = 4/4T

**Fig. 13.** Instantaneous fluid fields in terms of vorticity contour, streamline trace, and pressure distribution for Caudal Fin-like Shape A over one period of VIV oscillation cycle.

Fig. 13 shows the fluid field of Caudal Fin-like Shape A, in which the regular 2S mode of vortex shedding is also observed. Three separate points (i.e., points A, B, and C) exist in the flow. Vorticity occurs only after point C due to the obstructing effect of the fairing attachment. In other words, the shear layer separates from the tips of fairing fins instead of the cylinder surface to form vorticity, which delays the formation of vortex shedding. This phenomenon provides a desired redistribution of vorticity, which exerts a positive influence on the dynamic feature of the near wake. Similar to separate points in the plain cylinder, the present specific separate points also vary obviously depending on the location of the device. The color shade in Fig. 13 is lighter than that in Fig. 12, which implies that the magnitudes of vorticity and pressure of fluid field of Caudal Fin-like

Shape A are relatively small. This weakening of vorticity contributes to VIV suppression.

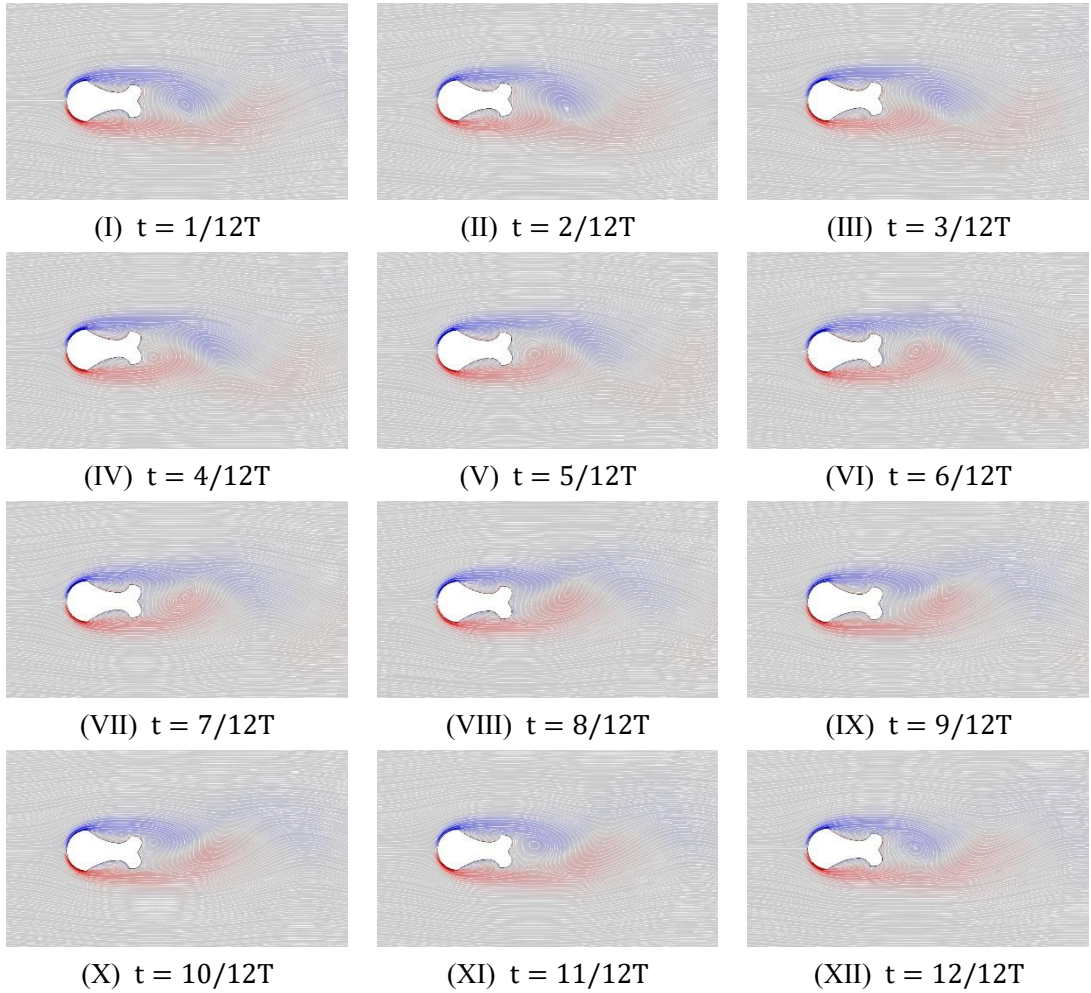

| (I) t = 1/12T | (II) t = 2/12T | (III) t = 3/12T |
| (IV) t = 4/12T | (V) t = 5/12T | (VI) t = 6/12T |
| (VII) t = 7/12T | (VIII) t = 8/12T | (IX) t = 9/12T |
| (X) t = 10/12T | (XI) t = 11/12T | (XII) t = 12/12T |

**Fig. 14.** Instantaneous fluid fields in terms of vorticity contour and streamline trace for Caudal Fin-like Shape B over one period of VIV oscillation cycle.

We evaluate Caudal Fin-like Shape B with a series of snapshots of the fluid field in terms of streamline traces and vorticity contours to further elucidate the reasons of the optimal fairing riser for VIV suppression. We select reduced velocity $Ur = 5.5$ to conduct the analysis due to the high performance of Caudal Fin-like Shape B for VIV reduction according to Fig. 9 and the above discussion. Fig. 14 shows that the magnitude of vorticity is relatively small to conduce to stabilize the near-wake region behind the body. The shear layer separates from the tips of fairing fins to delay the formation of vortex shedding, which conforms to the previous discussion about Caudal Fin-like Shape A. A pair of vorticities are generated at the bilateral of the optimal fairing before a vorticity form at the tips of fairing fins alternately. These two closed circulations with opposite directions are formed in the region between the cylinder and the fairing attachment, as shown in Fig. 14 (III, IV, IX, X), which maintain the fairing stability in flow. A similar phenomenon was described by Law et al. (2017) in their numerical studies of connected-C and disconnected-C devices.

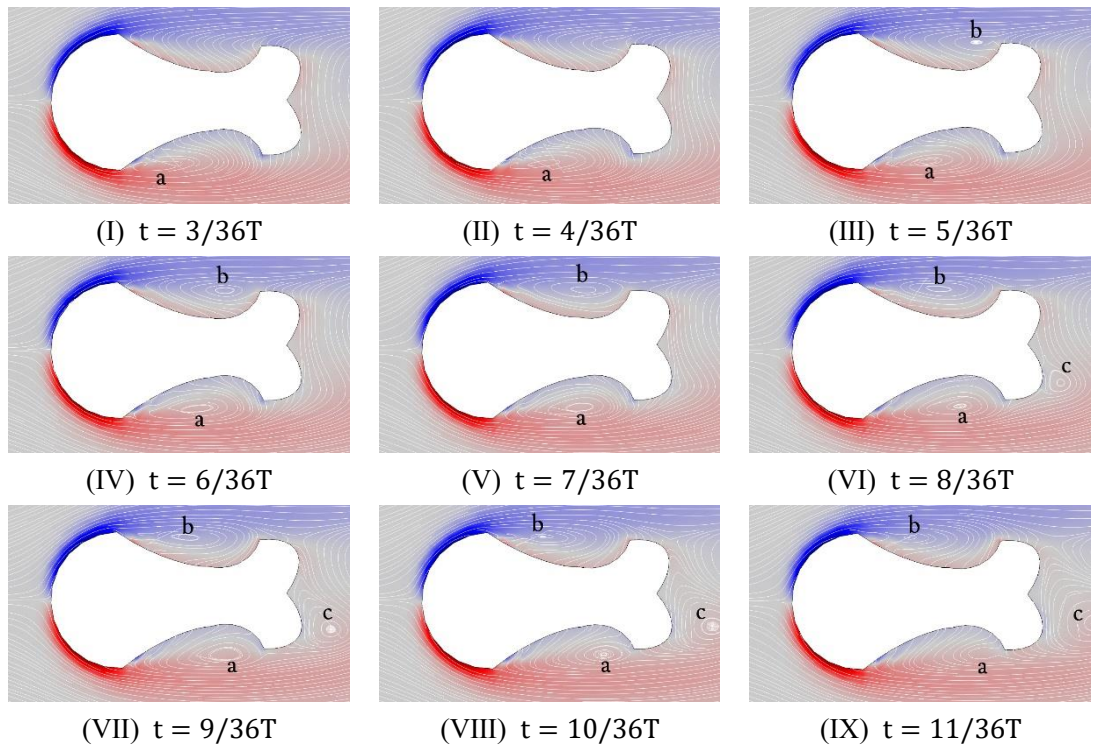

**Fig. 15.** Instantaneous fluid fields in terms of vorticity contour and streamline trace for Caudal Fin-like Shape B over a quarter period of VIV oscillation cycle.

The instant vorticity fields over one quarter oscillation period are detailed in Fig. 15 to describe the mechanism of the vorticity generation at the bilateral of the optimal fairing. Vorticity a forms near the bottom of the cylinder when $t = 3/36T$ in Fig. 15 (I), and vorticity b forms near the left of the tip of the fairing fin when $t = 5/36T$ in Fig. 15 (III). These two closed circulations with opposite directions move to left or right. When $t = 7/36T$–$8/36T$, these two vorticities are symmetric along the fairing streamwise, which conduce to maintain the fairing stability in the flow. Thus, the fairing meets the balance point in terms of amplitude when $t = 9/36T$. In Fig. 15 (VI), another vorticity (i.e., vorticity c) forms at the right of the tip of the fairing fin when $t = 9/36T$. Vorticities a and b then break up and disappear until $t = 11/36T$.

To sum up, two major factors contribute to the VIV suppression of the aforementioned devices according to the above analysis of vortex pattern. The first one is the location of the separate point that generates vorticity. Caudal Fin-like Shapes A and B verify the well-known phenomenon that the extension of the shear layer from the cylinder body to farther into the wake not only delays the formation of vorticities but also decreases the magnitude of vorticities. The second one is the formation of a pair of vorticities with opposite directions generated at the bilateral of the optimal fairing. Such two symmetric closed circulations conduce to maintain the fairing stability in the flow. This mechanism is observed in the wake structure of Caudal Fin-like Shape B.

## 5. Conclusions

Shape optimization of a two-dimensional riser fairing is conducted by integrating CFD and GA. The fairing profiles are parameterized by using Bézier curves, and the shape optimization is conducted at laminar flow with Reynolds number $Re \in [100,400]$. We implement two assessments with three cases (i.e., Cases A, B, and C) to analyze and validate the effectiveness of shape optimization.

In the first assessment with Cases A and B, the fairings are allowed to move in transverse and streamwise directions (i.e., 2-DOF motion). Fluid velocity plays a more significant role in shape optimization than the number of design parameters. The VIV response shows that all optimal shapes have high performances in VIV suppression in terms of transverse amplitude and lift and drag coefficients. This result may explain the phenomenon that the effect of the number of parameters is minimal. The limitation of 2-DOF prevents the requirement to evolve a new shape that is far from the water drop shape. A water-drop-shaped fairing exhibits a good performance of VIV suppression.

In the second assessment with Case C, the fairings are free to rotate and are allowed to move in transverse and streamwise directions (i.e., 3-DOF motion). Two interesting optimal caudal fin-like shapes are acquired. All the optimal shapes have good performances of VIV suppression at some range of fluid velocity. In fact, the optimal fairing starts to generate vigorous vibration excited by galloping when fluid velocity exceeds a specific velocity. The flow field in terms of instantaneous vorticities, streamlines, and pressure fields illustrates two factors that conduce to stabilize the wake region of fairing. The first one is the extension of the shear layer from the cylinder body to farther into the wake to delay the formation of vorticities, thereby decreasing the magnitude of vorticities. The second one is the formation of two symmetric vorticities with opposite directions generated at the bilateral of the optimal fairing, which maintain the fairing stability in the flow.

Further research is required toward a condition that conforms to the physical environment. However, shape optimization in the present work is limited to conduct in the condition of laminar flow with low Reynolds number considering the massive computational consumption. In fact, the calculation process of shape optimization in this study lasts for several days or more through a workstation with Intel Xeon E5. Thus, before beginning further research in turbulence flow, we need to solve the problem of massive computational consumption of shape optimization by two methods. First, a fast approach that simulates the vibration of structures in turbulence flow is required. Second, an efficient algorithm that conducts optimization is necessary. The proposed optimal shape of VIV suppressor should also be testified through a physical experiment.

## Acknowledgments

The authors wish to acknowledge the financial support of the National Program on Key Basic Research Project (2015CB251203), Program for Changjiang Scholars and Innovative Research Team in University (IRT14R58), Natural Science Foundation of Shandong Province (ZR2014EL018), Major National Science and Technology Program (2016ZX05028-001-05), the Fundamental Research Funds for the Central Universities (17CX02025A), and the Research Initiation Funds of China University of Petroleum (Y1703008).